\documentclass[doublecol]{epl2} 
\usepackage{amsmath}

\DeclareMathAlphabet{\mathpzc}{OT1}{pzc}{m}{it}
\newcommand{\m}{\mathpzc{m}}
\title{Hawking radiation in a two-component Bose-Einstein condensate}
\shorttitle{Hawking radiation in a two-component BEC} 

\author{P.-\'E. Larr\'e\inst{1} \and N. Pavloff\inst{1}}
\shortauthor{P.-\'E. Larr\'e \etal}

\institute{ \inst{1} Univ. Paris Sud, CNRS, L.P.T.M.S., UMR8626 - 91405 Orsay,
  France} 

\pacs{04.62.+v}{Quantum fields in curved spacetime}
\pacs{03.75.Mn}{Multicomponent condensates}
\pacs{04.70.Dy}{Quantum aspects of black holes}

\abstract{
We consider a simple realization of an event horizon in the
  flow of a one-dimensional two-component Bose-Einstein
  condensate. Such a condensate has two types of quasiparticles; In
  the system we study, one corresponds to density fluctuations and the
  other to polarization fluctuations. We treat the case in which a
  horizon occurs only for one type of quasiparticles (the
  polarization ones). We study the one- and two-body signal associated
  to the analog of spontaneous Hawking radiation and demonstrate by
  explicit computation that it consists only in the emission of
  polarization waves. We discuss the experimental consequences of the
  present results in the domain of atomic Bose-Einstein condensates and
  also for the physics of exciton-polaritons in semiconductor microcavities.}

\begin{document}

\maketitle

An intense research activity has been developed in the recent years aiming at
identifying Hawking radiation in several analog models of gravity
(see refs.~\cite{Bar11,Rob12} for recent reviews). 
The possible black hole configurations realized in an analogous system
all rely on the remark by Unruh \cite{Unr81} that if the flow of a
fluid has, while remaining stationary, a transition from a
subsonic upstream region to a supersonic downstream region, the
interface between these two regions behaves as an event horizon for
sound waves. The supersonic region mimics the interior of a black hole
since no sound can escape from it (one speaks of ``dumb hole'').  This
analogy is richer than a mere realization of a sonic event horizon:
Quantum virtual particles can tunnel out near the horizon and are then
separated by the background flow giving rise to correlated currents
emitted away from the region of the horizon (both inside and outside
of the black hole), in exact correspondence with the original scenario of
Hawking radiation \cite{Haw75}.

Among the prominent experimental configurations where a sonic horizon
has been realized one can quote the use of ultrashort pulses moving in
optical fibers \cite{Phil08} or in a dielectric medium \cite{Bel10},
the study of the flow of a Bose-Einstein condensate (BEC) past an
obstacle \cite{Lah10}, of a laser propagating in a nonlinear luminous
liquid \cite{Ela12}, or of surface waves on moving water
\cite{Rou08,Wei11}. Several recent theoretical works proposed other
realizations of an artificial event horizon, using
for instance an electromagnetic wave guide \cite{Sch05} (or more
recently a SQUID array transmission line \cite{Nat09}), ring-shaped
chain of trapped ions \cite{Hor10}, graphene \cite{Ior12,Che12}, or
edge modes of the filling fraction $\nu=1$ quantum Hall system
\cite{Sto13}. Among these theoretical proposals, those employing an
exciton-polariton superfluid \cite{Sol11,Ger12} deserve special
attention because they could be realized in a near future. Such
systems are specific because polaritons have an effective spin $1/2$
and, as we will see below, this has important qualitative consequences
on the expected Hawking signal.

In the present work we study the possible signatures of Hawking
radiation in a generic two-component BEC system. Such a system is
peculiar in the sense that it sustains two types of elementary
excitations, with different long-wavelength velocities.  This makes it
possible to realize a unique configuration where an event horizon
occurs for one type of excitations but not for the other.  The
associated artificial black hole could be experimentally implemented
in a polariton condensate (such as proposed in ref.~\cite{Ger12}), but
also in a two-species BEC such as realized by considering for instance
$^{87}$Rb in two hyperfine states \cite{Hal98}, or a mixture of two
elements \cite{Mod02}, or different isotopes of the same atom
\cite{Pap08}. 
A general theory of such
systems requires to consider a wide range of parameters and of
different situations corresponding to possibly different masses of the
two species, to different strengths and signs of intra- and
inter-species interactions, to different types of external potentials
(possibly species-dependent) and of coupling between the two
components. In the present work we consider a simple model which
captures the essential physical ingredients and characteristics of the
phenomenon: The order parameter of the two-component BEC is described
by a one-dimensional (1D) two-component Heisenberg field operator
$(\hat\psi_+(x,t),\hat\psi_-(x,t))$ obeying a set of coupled
Gross-Pitaevskii equations:
\begin{align}\label{e1}
\notag {\rm i}\hbar\, \partial_t\hat\psi_{\pm} 
= &-\frac{\hbar^2}{2 m}\,\partial_x^2 \hat\psi_{\pm}
+U(x)\,\hat\psi_{\pm} \\
&+\big[g_1 \, \hat{n}_{\pm} + g_2(x)\, \hat{n}_{\mp}\big]\,\hat\psi_{\pm}
-\mu\,\hat\psi_{\pm} \, .
\end{align}
In this equation $\hat{n}_{\pm}(x,t)=\hat\psi^\dagger_{\pm}(x,t)\,
\hat\psi_{\pm}(x,t)$ is the density of the $(\pm)$-component, $U(x)$
is an external potential, $\mu$ is the chemical potential, and
$g_1$ ($g_2$) is the intra-species (inter-species) contact-interaction
coupling constant. We choose to work
in a configuration where $0<g_2<g_1$. This is quite realistic for
atomic condensates (provided one neglects the small difference of the
interaction constant between $+/+$ and $-/-$ components). For
excitonic polaritons it is accepted that $g_1>0$ and that $|g_2|<g_1$,
in agreement with the observed overall repulsion between polaritons,
but it is typically believed that $g_2<0$. However, depending on the
detuning between the photon and the exciton modes (and on the proximity
with the bi-exciton resonance), $g_2$ may be positive or negative, as
observed in refs.~\cite{Vla10,Para10}. Our choice to consider the case
of a positive $g_2$ parameter will make it possible to treat a setting
where the event horizon occurs for a flow velocity inferior to the one
of ordinary sound.

We consider an idealized model in which $g_2$ and the external
potential $U$ both depend on $x$ in a way that ensures the existence
of a homogeneous and stationary classical solution of eq.~(\ref{e1})
of the form
\begin{equation}\label{e2}
\Psi_{\pm}(x)=\sqrt{n_0/2} \, \exp({\rm i}k_0 x) \, .
\end{equation}
This can be realized by considering a step-like configuration for which
$U(x) = U_u\,\Theta(-x) + U_d\,\Theta(x)$ and
$g_2(x) = g_{2,u}\,\Theta(-x) + g_{2,d}\,\Theta(x)$
(where $\Theta$ is the Heaviside step function) with
\begin{equation}\label{e3}
U_u+g_{2,u}\,n_0/2=U_d+g_{2,d}\,n_0/2.
\end{equation}
\begin{figure}
\includegraphics[width=0.95\linewidth]{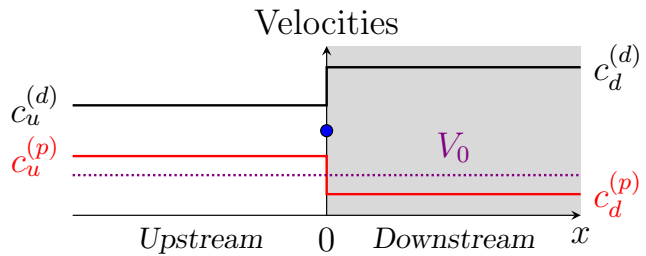}
\caption{(Colour on-line) Velocities of the ordinary sound $c^{(d)}$
  (black solid line) and of the polarization sound $c^{(p)}$ (red
  solid line) as a function of $x$. The dotted horizontal line
  represents the constant velocity $V_0$ of the flow. The blue dot is
  the reference velocity $\sqrt{g_1 n_0/(2 m)}$. The downstream region
  is shaded in order to recall that it corresponds to the interior of
  the black hole.}
\label{fig.1}
\end{figure}
The order parameter (\ref{e2}) describes a uniform flow in which both
components have the same density $n_0/2$ and the same velocity
$V_0=\hbar k_0/m$. We consider the case $V_0>0$, and denote the $x<0$
($x>0$) region as the upstream (downstream) region. In each of these
regions the long-wavelength elementary excitations consist either in
density or in polarization fluctuations with respective velocities
denoted as $c_\alpha^{(d)}$ and $c_\alpha^{(p)}$ ($\alpha=u$ or $d$,
depending if one considers the upstream or the downstream
region). $c_\alpha^{(d)}$ is the usual speed of sound whereas
$c_\alpha^{(p)}$ will be termed ``polarization sound velocity''; Their
precise definition will be given later [after Eqs. (\ref{e6}) and
(\ref{e7})].  As illustrated in fig. \ref{fig.1} we choose the
parameters of the system in such a way that
\begin{equation}\label{e4}
c_d^{(p)}<V_0<c_u^{(p)}<c_u^{(d)}<c_d^{(d)}\, .
\end{equation}
Then the point $x=0$ is an event horizon for the
fluctuations of polarization but not for the fluctuations of density
(the usual sound).

Note that the configuration we consider is of the same type as the one
considered in refs.~\cite{Bal08,Car08} for a one-component system, and
seems rather awkward: It consists in a uniform flow of a 1D BEC in
which the two-body interaction varies spatially (in order to locally
modify the speed of polarization sound in the system) although the
velocity and the density of the flow remain constant. This is only
possible in the presence of an external potential specially tailored
so that the local chemical potential remains constant everywhere [this
is ensured by eq.~(\ref{e3})]. This makes the whole system quite
difficult to realize experimentally. However, it was shown in
refs.~\cite{Lar12} and \cite{Ger12} that the Hawking radiation
associated to this configuration has the same properties as others
associated to more realistic realizations of an event horizon in a BEC
or a polariton condensate.

\begin{figure}
\begin{center}
\includegraphics[width=0.95\linewidth]{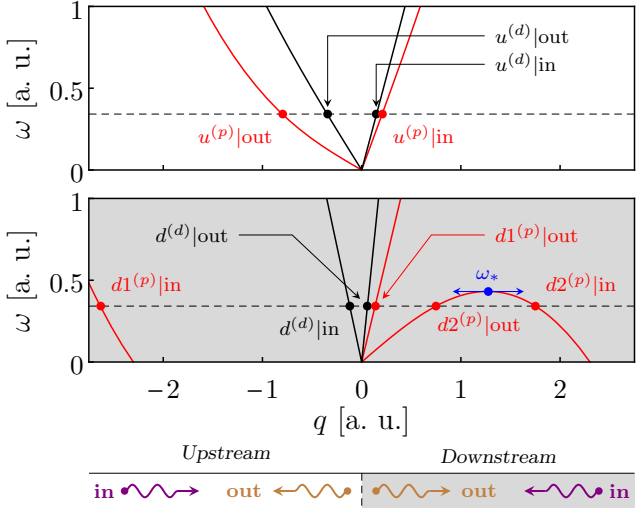}
\end{center}
\caption{(Colour on-line) Upstream (upper plot) and downstream (middle
  plot) dispersion relations. The red (black) curves correspond to
  polarization (density) modes.  In each plot the horizontal dashed
  line is fixed by the chosen value of $\omega$. The labeling of the
  different branches is explained in the text. The abscissae of the
  dots fix the values of the wavevector $q(\omega)$ corresponding to
  each branch. The lower diagram illustrates the terminology used in
  the main text for denoting the waves as outgoing or ingoing. As in
  fig.~\ref{fig.1}, the downstream region is shaded in order to recall
  that it corresponds to the interior of the black hole.}
\label{fig.2}
\end{figure}

The black hole configuration being fixed, we now characterize the
spontaneous Hawking emission by studying the quantum Bogoliubov
excitations of the system, in a manner similar to what has been done
in refs.~\cite{Leo03,Mac09,Rec09}. The most efficient way to
characterize the different branches of the dispersion relation is to
consider the classical (or more precisely first quantized) version of
eq.~(\ref{e1}). One writes the order parameter as $\psi_\pm(x,t) =
\Psi_\pm(x) + \delta\psi_\pm (x,t)$ with $|\delta\psi_{\pm}|\ll
|\Psi_\pm|$. In a region where $U(x)$ and $g_2(x)$ have the constant
value $U_{\alpha}$ and $g_{2,\alpha}$ ($\alpha=u$ or $d$) the
fluctuations with given pulsation $\omega$ on top of the background
(\ref{e2}) are of the form
\begin{equation}\label{e5}
\delta\psi_{\pm}(x,t)=\mathrm{e}^{{\rm i}k_0 x}
\big[u_{\pm,\alpha}(x,\omega)\,\mathrm{e}^{-{\rm i}\omega t}
+
w_{\pm,\alpha}^*(x,\omega)\,\mathrm{e}^{{\rm i}\omega t}\big] \, ,
\end{equation}
where the $u_{\pm,\alpha}$'s and the $w_{\pm,\alpha}$'s are plane
waves of momentum $\hbar \, q$. The corresponding dispersion relations are
represented in fig.~\ref{fig.2}.  The curves corresponding to density
fluctuations are represented in black in the figure. Their dispersion
relation reads
\begin{equation}\label{e6}
(\omega-V_0 \, q)^2 = [c_\alpha^{(d)}]^2 q^2
+ \frac{\hbar^2 q^4}{4 m^2} \, ,
\end{equation}
with $m [c_\alpha^{(d)}]^2 =\frac{1}{2}(g_1+g_{2,\alpha})n_0$
($\alpha=u$ or $d$). In the left-hand side of eq.~(\ref{e6}) the term
$-V_0 \, q$ is a Doppler shift indicating that the dispersion relation
is evaluated in the laboratory frame in which the flow has a constant
and uniform velocity $V_0$. In the upstream region the different
channels corresponding to (\ref{e6}) are denoted as $u^{(d)}|{\rm in}$
and $u^{(d)}|{\rm out}$, where the $u$ stands for ``upstream'', the
$(d)$ for ``density'' and the ``in'' (the ``out'') labels the wave
whose group velocity is directed towards (away from) the horizon. What
is considered as an ingoing or an outgoing wave is pictorially
represented in the lower diagram of fig.~\ref{fig.2}. In the
downstream region the channels are accordingly denoted as
$d^{(d)}|{\rm in}$ and $d^{(d)}|{\rm out}$ (see fig.~\ref{fig.2}).

The curves corresponding to fluctuations of the polarization
[$\pi(x,t) =n_+(x,t)-n_-(x,t)$] are represented in red in figure
\ref{fig.2}. Their dispersion relation is
\begin{equation}\label{e7}
(\omega-V_0 \, q)^2 = [c_\alpha^{(p)}]^2 q^2
+ \frac{\hbar^2 q^4}{4 m^2} \, ,
\end{equation}
with $m [c_\alpha^{(p)}]^2 =\frac{1}{2}(g_1-g_{2,\alpha})n_0$. In the upstream
region the corresponding channels are denoted as $u^{(p)}|{\rm in}$
and $u^{(p)}|{\rm out}$. In the downstream region, $V_0>c_d^{(p)}$,
and new branches appear in the dispersion relation of polarization
waves. Altogether in this region the branches are denoted as
$d1^{(p)}|{\rm in}$, $d1^{(p)}|{\rm out}$, $d2^{(p)}|{\rm in}$ and
$d2^{(p)}|{\rm out}$ (see fig.~\ref{fig.2}). In the following we refer
to the quasiparticles corresponding to the dispersion relation
(\ref{e6}) as density quasiparticles and to those corresponding to
(\ref{e7}) as polarization quasiparticles.

The existence of the discontinuity in the parameters of the system at
$x=0$ prevents the channels we have just identified for an hypothetical
homogeneous configuration to be the true eigenmodes of the system. The
correct eigenmodes are linear combinations of the channels in the
upstream region and channels in the downstream one, with
appropriate matching at $x=0$. Among all the possible combinations, we
are primarily interested in the scattering modes which describe a
plane-wave excitation originating from infinity -- either upstream or
downstream -- on a well defined ingoing channel, impinging on the
horizon, and then leaving again towards infinity as a superposition of
the outgoing branches.   
When $\omega$ is lower than the threshold $\omega_*$ identified in
fig.~\ref{fig.2}, there are 5 ingoing channels and 5 outgoing ones.
The corresponding scattering amplitudes form a $5\times 5$ $S$ matrix
which can be shown to be block diagonal:
\begin{equation}\label{e8}
S= \left(
\begin{array}{c|c}
S^{(p,p)} & \begin{array}{cc}
0 & 0\\
0 & 0\\
0 & 0\end{array}
\\ \hline
\begin{array}{ccc}
0 & 0 & 0\\
0 & 0 & 0\end{array}
& S^{(d,d)}
\end{array}
\right)
,
\end{equation}
with
\begin{subequations}
\label{e9}
\begin{align}
\label{e9a}
&S^{(p,p)}=\left(
\begin{array}{ccc}
S_{u^{(p)},u^{(p)}} & S_{u^{(p)},d1^{(p)}} & S_{u^{(p)},d2^{(p)}} \\
S_{d1^{(p)},u^{(p)}} & S_{d1^{(p)},d1^{(p)}} & S_{d1^{(p)},d2^{(p)}} \\
S_{d2^{(p)},u^{(p)}} & S_{d2^{(p)},d1^{(p)}} & S_{d2^{(p)},d2^{(p)}}
\end{array}\right) , \\
\label{e9b}
& S^{(d,d)}=\left(
\begin{array}{cc}
S_{u^{(d)},u^{(d)}} & S_{u^{(d)},d^{(d)}} \\
S_{d^{(d)},u^{(d)}} & S_{d^{(d)},d^{(d)}} \\
\end{array}
\right) .
\end{align} 
\end{subequations}
For instance the $S_{u^{(p)},d1^{(p)}}$ matrix element denotes the
(complex and $\omega$-dependent) scattering coefficient from the
ingoing downstream channel $d1^{(p)}|{\rm in}$ towards the outgoing
upstream channel $u^{(p)}|{\rm out}$. As discussed in
refs.~\cite{Rec09,Mac09,Lar12}, current conservation imposes a skew
unitarity of the $S$ matrix: $S^\dagger \eta S=\eta$, where
here $\eta={\rm diag}(1,1,-1,1,1)$. When $\omega$ is larger
than the maximum $\omega_*$ of the $d2^{(p)}$ branches (see
fig.~\ref{fig.2}) the $d2|{\rm in}$ and $d2|{\rm out}$ channels
disappear, the $S^{(p,p)}$ submatrix becomes $2\times 2$, and the now
$4\times 4$ $S$ matrix obeys the usual unitarity condition $S^\dagger
S = {\rm diag}(1,1,1,1)$.

We computed the coefficients of the $S$ matrix both analytically (in
the low-$\omega$ limit) and numerically (for unrestricted values of
$\omega$). We checked the excellent agreement
between the two approaches in their common range of validity (i.e., at
$\omega\to 0$) and also that the current conservation conditions are
verified, exactly in the analytical approach, and with a high degree
of accuracy in the numerical treatment (the error is always less than
$10^{-7}$). All the matrix coefficients of the form $S_{i,d1^{(p)}}$
and $S_{i,d2^{(p)}}$ with $i \in \{ u^{(p)} , d1^{(p)} , d2^{(p)} \}$
(i.e., the two right most columns of $S^{(p,p)}$) diverge at low
$\omega$. This is connected to the fact that the associated Wigner
time delay diverges: Low-energy polarization quasiparticles entering
the system {\sl via} the $d1^{(p)}|{\rm in}$ or the $d2^{(p)}|{\rm
  in}$ channels -- i.e., from the interior of the black hole -- remain
blocked at the horizon forever: This is a signature of the occurrence
of an event horizon for the polarization modes. On the contrary,
low-energy density quasiparticles entering the system from the
downstream region can escape the black hole, since we work in a
configuration where the horizon does not affect the density
fluctuations (see fig.~\ref{fig.1}). Of course all quasiparticles
entering the system from the upstream region can cross the horizon and
penetrate into the black hole.

Within the present Bogoliubov analysis, the knowledge of the
$S$ matrix of the system makes it possible to characterize the
Hawking signal which corresponds to emission of radiation from the
interior toward the exterior of the black hole. In our specific case
the energy current associated to emission of elementary excitations is
(cf.~\cite{Kag03})
\begin{equation}\label{c11}
Q(x,t) = -\frac{\hbar^2}{2m} 
\sum_{\sigma=\pm 1}\left\langle \partial_t\hat\psi_\sigma ^\dagger(x,t) 
\, \partial_x\hat\psi_\sigma(x,t)\right\rangle  + {\rm H.c.}\, ,
\end{equation}
where ``H.c.'' stands for ``Hermitian conjugate''. $Q(x,t)$ is here
time and position-independent in agreement with the conservation of
the energy flux in a stationary configuration. Computing its
expression far upstream ($x\to -\infty$) one can show, as expected,
that the current is only carried by the $u^{(p)}|{\rm out}$ channel
and is, at zero temperature, given by the formula
\begin{equation}\label{c2}
Q=- \int_0^{\omega_*} \frac{{\rm d}\omega}{2\pi} \, \hbar\omega \,
\big|S_{u^{(p)},d2^{(p)}}(\omega)\big|^2 \, .
\end{equation}
Hence the quantity $|S_{u^{(p)},d2^{(p)}}(\omega)|^2$ characterizes
the emission spectrum of Hawking radiation. Although we consider a
setting with step-like variations of the external parameters,
resulting in an infinite effective surface gravity, the Hawking
spectrum is still thermal like, i.e., approximately of the form
\begin{equation}\label{c3}
\big|S_{u^{(p)},d2^{(p)}}(\omega)\big|^2
\simeq \frac{\Gamma}
{\exp\big(\frac{\hbar\, \omega}{k_{\rm B} T_{\rm H}}\big)-1}\, ,
\end{equation}
where $k_{\rm B}$ is the Boltzmann constant, $\Gamma$ is denoted
as the gray-body factor and $T_{\rm H}$ is the Hawking
temperature. Since we have computed the explicit low-$\omega$
expression of the coefficients of the $S$ matrix, we can determine
$T_{\rm H}$ and $\Gamma$ by a low-$\omega$ fit of expression
(\ref{c3}). In particular one obtains the following explicit
expressions for the reduced Hawking temperature ${\mathcal T}_{\rm H}=
k_{\rm B}T_{\rm H}/m [c^{(p)}_u]^2$ and for the gray-body factor:
\begin{equation}\label{c4}
{\mathcal T}_{\rm H}=
\displaystyle{\frac{1}{2} \frac{\m_u^2}{\m_d}
\frac{(1-\m_u^2)(\m_d^2-1)^{\frac{3}{2}}}{\m_d^2-\m_u^2}} \, ,
\quad
\Gamma=\frac{4 \m_u}{(1+\m_u)^2} \, ,
\end{equation}
where $\m_\alpha=V_0/c^{(p)}_\alpha$ is the (polarization) Mach number
in region $\alpha$ ($\alpha=u$ or $d$ and $\m_u<1<\m_d$). 

\begin{figure}
\includegraphics*[width=0.95\linewidth]{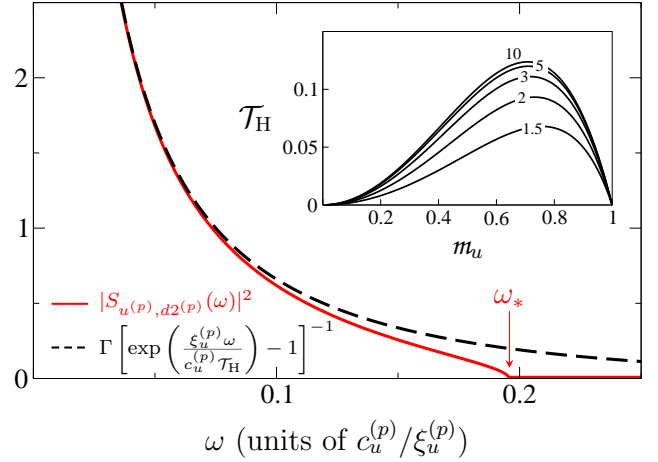}
\caption{(Colour on-line) $|S_{u^{(p)},d2^{(p)}}|^2$ (red solid line) and its
  approximation by eq.~(\ref{c3}) (black dashed line) as a function of
  $\omega$. We consider the case with
 $\m_u=0.7$ and $\m_d=3$. With the present choice of parameters,
  $\Gamma\simeq 0.969$, ${\mathcal T}_{\rm H}\simeq 0.111$ and $\omega_*
  \xi_u^{(p)}/c_u^{(p)}\simeq 0.196$, where $\xi_u^{(p)}=\hbar/(m
  c_u^{(p)})$. The inset displays ${\mathcal T}_{\rm H}$ as a function of
  $\m_u$ for several values of $\m_d$ ($\m_d=1.5, \hdots ,10$).}
\label{fig.onebody}
\end{figure}

The numerically determined $|S_{u^{(p)},d2^{(p)}}|^2$ is compared in
fig.~(\ref{fig.onebody}) with the thermal spectrum (\ref{c3}) where
$T_{\rm H}$ is given by (\ref{c4}). The plot is done in a
configuration where $\m_u=0.7$, $\m_d=3$ and  $g_{2,u}/g_1=0.2$. In
the type of setting we consider, fixing these three parameters determines
all the other relevant quantities of the system. In particular one has
here $g_{2,d}/g_1\simeq 0.956$, $V_0/c^{(d)}_u\simeq 0.572$ and
$V_0/c^{(d)}_d\simeq 0.448$. As expected one sees in the figure that the
(numerically) exact spectral density $|S_{u^{(p)},d2^{(p)}}|^2$
coincides with a thermal gray-body emission at low energy.  Note
however that $|S_{u^{(p)},d2^{(p)}}|^2$ is strictly zero for
$\omega>\omega_*$ since above this threshold the $d2^{(p)}|{\rm in}$
and $d2^{(p)}|{\rm out}$ channels disappear and the $S$ matrix becomes
$4\times 4$.

Formulas (\ref{c4}) show that the Hawking temperature is roughly of
order of $m [c^{(p)}_u]^2$, which itself is of order of the chemical
potential of the system. In atomic condensates the chemical potential
is of order of the temperature of the system and the Hawking current
will be hidden by the thermal noise. In polariton systems the
chemical potential is typically of order of $0.5$ meV and low
temperature experiments could in principle distinguish the Hawking
current from the thermal noise.

We now consider an other experimental observable which can reveal the
Hawking phenomenon even in the presence of a realistic thermal noise. As
first explicitly pointed out in refs.~\cite{Bal08,Car08}, in analog
systems an external observer is able to measure correlations across
the horizon revealing the existence of the Hawking current (see also
refs.~\cite{Mac09,Rec09,Sch10,Par10,Lar12}). For the present setting
we expect that these correlations are due to pair-wise emission of
polarization quasiparticles on both sides of the horizon. The
polarization density operator in our system is $\hat{\pi}(x,t) =
\hat{n}_+(x,t)- \hat{n}_-(x,t)$. In the configuration we consider it
has zero mean [$\langle\hat{\pi}(x,t)\rangle=0$] and the corresponding
correlation signal is time-independent:
\begin{equation}\label{c21}
\begin{split}
g^{(p)}(x,x')& = \langle :\! \hat{\pi}(x,t)\, \hat{\pi}(x',t)\! : \rangle \\
& = \langle \hat{\pi}(x,t)\, \hat{\pi}(x',t) \rangle -
\delta(x-x')\, n_0 \, .
\end{split}
\end{equation}
It is also interesting to study the correlation of the density fluctuations
\begin{equation}\label{c22}
\begin{split}
  g^{(d)}(x,x')& = 
\langle :\! \delta\hat{n}(x,t)\, \delta\hat{n}(x',t)\! : \rangle \\
  & = \langle \delta\hat{n}(x,t)\, \delta\hat{n}(x',t) \rangle -
  \delta(x-x') \, n_0 \, ,
\end{split}
\end{equation}
where $\delta\hat{n}(x,t) =\hat{n}_+(x,t) + \hat{n}_-(x,t)-n_0$.

In a hypothetical homogeneous configuration where $U(x)$ and $g_2(x)$
have constant uniform values, these correlator
read
\begin{equation}\label{c23}
g^{(d,p)}(x,x')= 
\frac{n_0}{\xi^{(d,p)}}\, F\left(\frac{|x-x'|}{\xi^{(d,p)}}\right) ,
\end{equation}
where $F(z) = - (\pi z)^{-1} 
\int_0^\infty {\rm d}t \, \sin(2 \, t \, z) \, (1+t^2)^{-3/2}$ and
$\xi^{(d,p)}=\hbar/m c^{(d,p)}$.

The correlation patterns (\ref{c23}) are drastically modified in the
presence of an event horizon. There is a first trivial modification
due to the space dependence of the speeds of sound: Formulas
(\ref{c23}) are modified upstream and downstream of the horizon
because, in the region $x<0$, the values of $\xi^{(d)}_u$ and
$\xi^{(p)}_u$ are different from those of $\xi^{(d)}_d$ and
$\xi^{(p)}_d$ in the region $x>0$. The second modification corresponds
to long-distance correlations and is more interesting: Quantum
fluctuations generate correlated currents of polarization
quasiparticles propagating away from the horizon in the $u^{(p)}|{\rm
  out}$, $d1^{(p)}|{\rm out}$ and $d2^{(p)}|{\rm out}$ channels. This,
in turn, induces long-range modifications of $g^{(p)}(x,x')$.  No such
long-distance correlations are expected for $g^{(d)}(x,x')$ since
there is no horizon for the density quasiparticles.

\begin{figure}
\centering
\includegraphics*[width=0.95\linewidth]{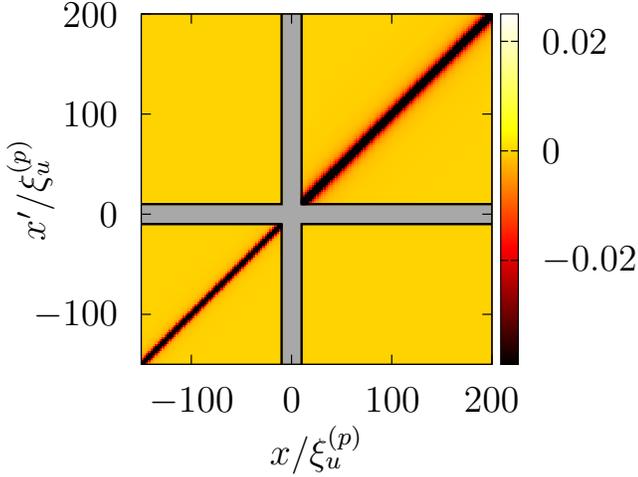}
\caption{(Colour on-line) 2D plot of the numerical result for the dimensionless
  quantity $\xi_u^{(p)} g^{(d)}(x,x')/n_0$ in the case in which
  $g_{2,u}/g_1=0.2$, $\m_u=0.7$ and $\m_d=3$. The shaded area near the
  axis corresponds to the zone $|x|$ or $|x'|<10\, \xi_u^{(p)}$.}
\label{fig.g2dens}
\end{figure}

\begin{figure}
\centering
\includegraphics*[width=0.95\linewidth]{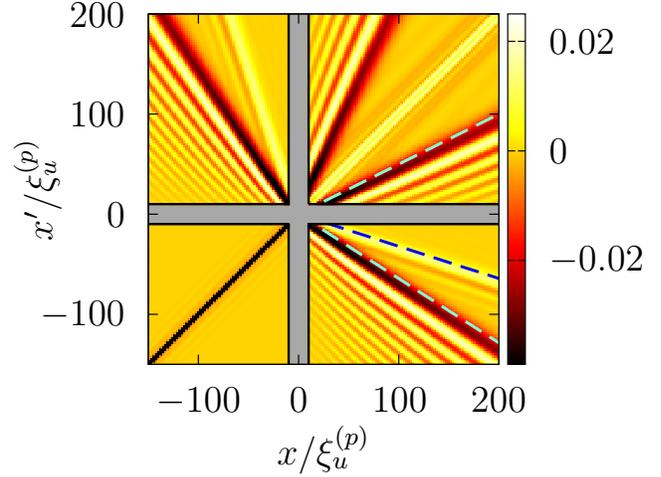}
\caption{(Colour on-line) Same as fig.~\ref{fig.g2dens} for the
  dimensionless quantity $\xi_u^{(p)} g^{(p)}(x,x')/n_0$. The
  dashed straight lines correspond to the correlation lines where a
  heuristic interpretation of the Hawking signal leads to expecting the
  largest long-range signal (see the text).}
\label{fig.g2pol}
\end{figure}

The knowledge of the $S$ matrix makes it possible to explicitly
compute the quantities $g^{(p)}(x,x')$ and $g^{(d)}(x,x')$ if $x$ and
$x'$ are not too close to the horizon\footnote{In vicinity of the
  horizon one should take into account evanescent modes for accurately
  evaluating the correlation signal. This makes the computation
  cumbersome although poorly instructive. This is the reason why in
  figs.~\ref{fig.g2dens} and \ref{fig.g2pol} we exclude the regions
  where $|x|$ or $|x'|$ are lower than $10\, \xi_u^{(p)}$.}. We do not
write down here the extensive explicit formulas (see
ref.~\cite{Lar13}) but rather display a plot of both quantities
$g^{(d)}(x,x')$ (fig.~\ref{fig.g2dens}) and $g^{(p)}(x,x')$
(fig.~\ref{fig.g2pol}) in a black hole configuration with
$g_{2,u}/g_1=0.2$, $\m_u=0.7$ and $\m_d=3$ (the same parameters for
which fig.~\ref{fig.onebody} has been drawn). As expected no track of
Hawking radiation can be observed in the plot of $g^{(d)}(x,x')$. On
the other hand, $g^{(p)}(x,x')$ displays long-range correlations along
three special directions highlighted in fig.~\ref{fig.g2pol} by dashed
straight lines.  According to the standard scenario of Hawking
radiation\cite{Haw75}, if correlated low-energy Hawking quasiparticles
are emitted along the $u^{(p)}|{\rm out}$, $d1^{(p)}|{\rm out}$ and
$d2^{(p)}|{\rm out}$ channels, at time $t$ after their emission, these
phonons are respectively located at positions $(V_0-c_u^{(p)})\,t <0$
\footnote{As clear from eq.~(\ref{e7}) and fig.~\ref{fig.2},
  $V_0-c_u^{(p)}$ is the $\omega\to 0$ limit of the group velocity of
  outgoing upstream polarization quasiparticles.}, $
(V_0+c_d^{(d)})\,t >0$, and $(V_0-c_d^{(p)})\,t >0$. This induces a
correlation signal along the lines of slopes
$(V_0-c_u^{(p)})/(V_0+c_d^{(p)})$ (resulting from correlations between
phonons emitted along the $u^{(p)}$ and $d1^{(p)}$ outgoing channels),
$(V_0-c_u^{(p)})/(V_0-c_d^{(p)})$ ($u^{(p)}|{\rm out}-d2^{(p)}|{\rm
  out}$ correlations), and $(V_0-c_d^{(p)})/(V_0+c_d^{(p)})$
($d2^{(p)}|{\rm out}-d1^{(p)}|{\rm out}$ correlations). These are the
three slopes marked by dashed lines in fig.~\ref{fig.g2pol}. These
large-distance correlation lines are accompanied by diffractive
corrections building an oscillatory pattern in their vicinity (see,
e.g., the discussion in ref.~ \cite{Rec09}). Of course the lines with
inverse slopes are also present (they correspond to the exchange $x
\leftrightarrow x'$ in fig.~\ref{fig.g2pol}).  The fact that, in the
present setting, such a pattern is observed in the correlation of
polarization fluctuations but not in the correlation of density
fluctuations is a strong demonstration that this signal is
intrinsically connected to Hawking radiation and requires the
occurrence of a horizon.

The experimental detection of the polarization signal described in the
present work is simple in the case of a polariton condensate because
the pseudo-spin of the decaying polaritons is commuted into right or left
circular polarization of the emitted photons. Also, the high repetition
rate achieved in this type of experiment should make it possible to
obtain a good statistics leading to a precise evaluation of the
correlation signal. For atomic condensates on the other hand, the
imaging techniques may rely on Stern-Gerlach and time-of-flight
analysis or dispersive optical measurements \cite{Car04} (for a
review, see, e.g., \cite{Sta13}).

Finally we note that the present treatment of vacuum fluctuations in a
stationary configuration, which is valid for a stable/conservative
atomic condensate, does not immediately apply for a nonequilibrium
polariton condensate. Indeed polaritons have a finite lifetime and the
vacuum fluctuations such as described in the present stationary
situation strictly speaking disappear, because no ingoing mode issued
from infinity is able to reach the horizon. The fluctuations of the
system are now triggered by fluctuations inside the excitonic
reservoir and by the losses. A related view concerns the dispersion
relation plotted in fig.~\ref{fig.2}: Because of damping, the
frequency of the normal modes typically acquire an imaginary part, and
long-wavelength density modes even become completely diffusive (see,
e.g., the review \cite{Car13}). However, one can show, within a simple
model of nonresonant pumping with gain and loss, that these damping
effects which are indeed present in the density channel, only weakly
affect the polarization mode \cite{Lar13b} and we thus expect that the
results of the present work should be also observable in future
experiments on out-of-equilibrium polariton condensates.

\acknowledgments It is a pleasure to thank A. Amo, I. Carusotto,
S. Finazzi and A. Recati for stimulating discussions. This
work was supported by the French ANR under grant n$^\circ$
ANR-11-IDEX-0003-02 (Inter-Labex grant QEAGE).

\end{document}